|
| Keywords | plasma agriculture; maize cultivation; dry atmospheric plasma priming; plasma activated water; chlorophyll content; stomatal density |

# 1. Introduction

Maize is the cereal with the highest production volume worldwide, surpassing even wheat and rice in terms of tonnage [1]. The United States (361 Mt/year) and China (259 Mt/year) dominate the maize production, followed by Brazil, Argentina, Ukraine and India [2]. Together, these 8 countries account for 881 Mt/year (*i.e.* 77.4%) of the global maize production. This cereal also plays a critical economic role in rural areas of many developing countries from Asia and Africa, where it is a key component of food security and income for smallholder farmers [3]. Hence in Cameroon, maize is not just a food staple for over 15 million people in the country, but also serves as the principal raw material for food processing, brewing, livestock feed and other agro-based industries [4]. Beyond the food aspect, this cereal can be utilized in industrial products, notably biofuels (especially ethanol), sweeteners and biodegradable plastics [5], [6], [7].

Global maize production faces several challenges including vulnerability to pests like the fall armyworm, which has spread to Asia and Africa, causing significant crop damage [8], [9]. Climate change poses another major challenge, affecting maize yields through altered rainfall patterns and increased temperatures. As underlined by Kim *et al.*, drought conditions can reduce maize yields significantly, with field experiment data showing a reduction of up to 39.3% in maize yields due to approximately 40% water loss [10]. The same study highlights that temperatures above 32°C, especially during the critical pollination and silk emergence period, negatively impact maize yields. In addition to these issues, it is also worth stressing that maize cultivation raises significant environmental concerns, including excessive water consumption [11], [12], land degradation [13] and the adverse effects of fertilizers and pesticides on ecosystems (chlorpyrifos, neonicotinoids, glyphosate) [14], [15].

To address these challenges, advancements in agricultural technology are proving crucial. Genetically Modified Organisms (GMOs) are being developed to resist pests like the fall armyworm [16], [17] but also to withstand climatic stressors [18]. This strategy aims at reducing reliance on chemical pesticides and enhancing drought resilience. As an alternative to GMOs, drought-tolerant maize varieties such as SAMMAZ-26 are also being developed to cope with less predictable rainfall and higher temperatures [19], [20]. Integrated Pest Management (IPM) strategies combine various agricultural practices to sustainably manage pest populations [21], [22]. Innovations in soil health, such as no-till farming and organic amendments, help mitigate land degradation [23], [24]. Nevertheless, these interventions alone may not be sufficient to meet the multiple challenges, hence the need to turn to emerging technologies such as cold plasma.

Cold plasma, or non-thermal plasma, is a technology that involves ionizing gases using electric discharges at low temperatures, creating a highly active environment where free electrons, ions and excited molecules coexist without significant heat generation [25]. In agriculture, it is primarily used for seed treatment, enhancing germination, root length, growth rate and pathogen decontamination [26], [27], [28]. This innovative approach operates without chemical inputs, making it an eco-friendly alternative that promotes plant health and resilience by modifying surface properties and microbial loads on seeds.

Based on the agricultural issues to be addressed, the research published so far has focused on the use of cold plasmas, either through dry or liquid processes. Dry plasma processes include sources operating at atmospheric pressure (dielectric barrier discharges, gliding arcs and plasma jets) or low pressure (DC discharges, microwaves and RF discharges), while all liquid plasma processes operate at atmospheric pressure [29]. Surprisingly, the combination of dry and liquid process methods has only been investigated by two teams to date. Hence, Sivachandiran *et al.* treated radish, tomato and sweet pepper seeds with DBD, followed by irrigating the seedlings with PAW produced by submerging a plasma jet into demineralized water [30]. Similarly, Rashid *et al.* treated paddy seeds using an AC glow discharge at low pressure and subsequently hydrated the seedlings with PAW generated by exposing demineralized water to a plasma jet [31]. Our builds upon this innovative approach, using maize as a field crop model and adapting a number of experimental conditions to study the synergy of the two following processes:
- Pre-germinative process: seeds priming by a dry atmospheric plasma (DAP) generated in a dielectric barrier device (DBD);
- Post-germinative process: seedlings irrigation by plasma-activated water (PAW) synthesized using a DC discharge generated by a single pin electrode device (SPED).

Conversely to the works of Sivachandiran *et al.* and Rashid *et al.*, our two plasma processes operate both at atmospheric pressure and seedlings are irrigated with plasma-activated tap water.

# 2. Materials & Methods

## 2.1. Plant material

Maize (*Zea mays* L.) seeds are produced during September 2023 at the Station Polyvalente de Recherche Agricole de Dschang (SPRAD), in Cameroon. The composite variety ATP-SR-Y is chosen because of its popularity with growers and its adaptability to the monomodal rainforest agro-ecological zone (Zone III). Seeds are stored in glass containers under laboratory conditions (18-22°C, 8-12% relative humidity).

## 2.2. Tap water

Tap water (TW) from the public network in Paris is collected during March, April and May of 2024. The composition of this water (**Table 1**) is measured by the autonomous public utility of the city of Paris (France), responsible for water supply and distribution.

*Table 1. Comparison of tap water composition from Paris public water supply and national quality standards (France) [32].*

| Water chemistry parameters | Concentration (mg/L) | |
|---|---|---|
| | Official measurements from "Eau de Paris" laboratory | National quality standards |
| Calcium | 90 | - |
| Magnesium | 6 | - |
| Sodium | 10 | 200 |
| Potassium | 2 | 12 |
| Bicarbonates | 220 | - |
| Sulfates | 30 | 250 |
| Chlorides | 20 | 250 |
| Nitrates | 29 | 50 |
| Fluoride | 0.17 | 1.5 |
| Total mineral content dry extract at 180 °C | 420 | - |

## 2.3. Seeds, seedlings and leaves monitoring

Germination is monitored by placing the seeds (50 seeds/group) on absorbent paper fully saturated with tap water, in a thermal enclosure at 15°C. The seeds are maintained without light to simulate a natural underground environment conducive to germination. For 15 days, the state of germination is observed and daily-documented. The onset of germination is defined when the seed radicle pierces the pericarp, as achieved in previous work [33].

A selection of 24 seedlings per condition is performed among the germinated seeds, 16 days after their imbibition. This selection is achieved by considering stem lengths of 2 mm (±1 mm). Once selected, the seedlings are sown in potting soil and irrigated either with tap water (TW) or plasma activated tap water (PAW) at days 16, 22, 26 and 33. Early stages of seedling growth is assessed by measuring following parameters: stem length, hypocotyl length with a graduate ruler accurate at 0.5 mm and collar diameter with a digital caliper (Dexter) accurate at 10 μm. The number of leaves is also quantified.

Chlorophyll content of leaves is measured using a chlorophyll meter (Spad-502PLUS, Konica Minolta). This portable device is designed for non-destructively measuring leaf chlorophyll content by clamping a leaf between its sensors to instantly assess plant health and nitrogen availability. Leaf chlorophyll concentration ($C_{chl}$) is expressed in $\mu mol.m^{-2}$ following Markwell's empirical formula: $C_{chl} = 10^{(M^{0.265})}$ where M is the value in arbitrary units returned by the SPAD 502PLUS device [34].

Dry and wet masses of seedlings (10 specimens per condition) are measured on an analytical balance (SARTORIUS, model Entris 124i-1S) with a readability of 0.1 mg and a reproducibility of ±0.1 mg.

The microstructure of maize leaves is analyzed on the 36th day using a fluorescence cell imaging device (Cytation 3 Cell Imager model from BioTek company). Observations are conducted at a 4× magnification under wavelengths typically used in DAPI mode (excitation at approximately 358 nm and emission around 461 nm). In this work, no DAPI dye agent is applied to the leaves, i.e. the observed fluorescence is due to the plant tissues themselves. This natural auto-fluorescence occurs when cellular and subcellular components inherent in the leaves absorb light at these specific wavelengths and re-emit it [35]. In our work, auto-fluorescence is utilized to specifically highlight the nuclei of guard cells within the stomata. Their auto-fluorescence is known to be more intense than that from the neighboring epidermal cells in a variety of both wild and cultivated plants [35].

## 2.4. Plasma experimental setups

DAP priming (**Fig. 1a**) is achieved by treating maize seeds by a cold plasma generated within a dielectric barrier device (DBD). This device consists of a rectangular quartz tube (30 cm in length, 1.5 mm in thickness, inner cross-section of 15 mm by 35 mm) lined by two 0.1 mm thick aluminum electrodes, 15 cm in length, hence designing a reactor volume of approximately 80 cm$^3$ where seeds can be placed. The DBD is powered by a high voltage generator that includes a function generator (ELC Annecy France, GF467AF) and a power amplifier (Crest Audio, 5500W, CC5500). This setup operates at an electrical input of 8 kV amplitude at 250 Hz (sine wave) and is supplied with a gas mixture of helium gas (1 slm) and air gas (0.5 slm), all maintained at atmospheric pressure. To guarantee a homogeneous plasma treatment of the seeds, the DBD is fixed onto a vibrating shaker (Heideloph company, Vibramax 100 model) which operates at 1250 rpm.

Plasma-activation of tap water (PAW) is obtained by placing 90 mL of tap water in a beaker (150 mL) and exposing it to a DC discharge generated in ambient air by the single pin electrode device (SPED). As shown in **Fig. 1b**, this device corresponds to a single hollow brass cylinder (10 cm in length, 2.4 mm inner and 3 mm outer diameters). It is inserted through the beaker's cover which isolates the plasma discharge and the PAW from the ambient environment. For all treatments, the SPED is powered by a DC high voltage generator (FuG company, HCN 350M-6500 model) supplying 4200 V and 30 mA. The liquid sample is swirled by a magnetic stirrer so that the resulting vortex is separated from the brass electrode by a gap of 5 mm. Finally, the liquid sample is electrically grounded by an electrical wire, as sketched in **Fig. 1b**. In this article, maize seedlings are irrigated either with PAW after 5 min of SPED exposure (PAW group) or with TW (Control group). The 5 min duration has been chosen as a good compromise between (i) generating sufficiently high levels of ROS to unambiguously observe biological effects (ii) while keeping in mind that the activation of water by plasma presents an energetical cost, especially in an outline of technological transfer.

## 2.5. Plasma phase characterization

Measurements of electrical parameters are carried out using a Teledyne LeCroy Wavesurfer 3054 digital oscilloscope, in conjunction with high-voltage probes (Tektronix P6015A 1000:1, Teledyne LeCroy PPE 20 kV 1000:1 and Teledyne LeCroy PP020 10:1) and a Pearson 2877 current monitor.

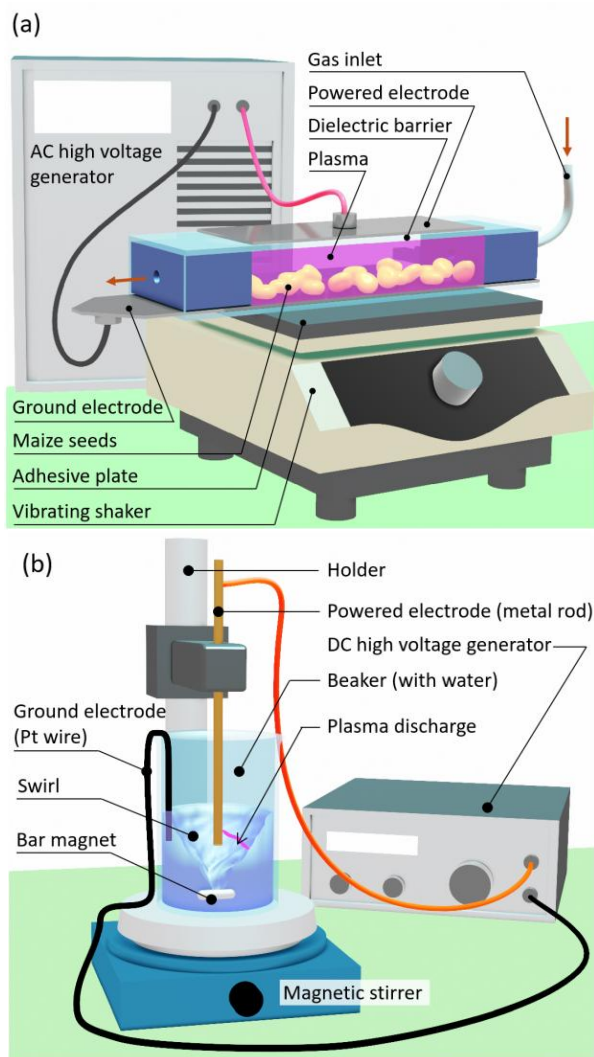

*Figure 1. Schematic diagram of (a) the dielectric barrier device (DBD) utilized for priming the maize seeds with cold plasma of He-air (b) the single pin electrode device (SPED) utilized for the activation of water by DC discharge in ambient air.*

The radiative components of the plasma phase are characterized by optical emission spectroscopy (OES), using an Andor SR-750-B1-R spectrometer equipped with an ICCD camera (model Istar) in a Czerny-Turner setup. The system has a focal length of 750 mm and a 1200 groove/mm grating, blazed at 300 nm. For all experiments, parameters include an exposure time of 0.1 s, 20 accumulations and a gain level of 2000.

## 2.6. Plasma-activated water characterization

Electrical conductivity is assessed in 20 mL of TW or PAW with a conductometer (HI-87314, Hannah) calibrated at 25°C on the 199.9-1999 µS/cm range with an accuracy of 1% (full scale). This device is calibrated before any measurement using distilled water (7.5 ± 0.5 µS/cm). As well as conductivity depending on temperature, PAW measurements are performed after water cooling to 25°C. pH, hardness, alkalinity and chemical elements (carbonates, cyanuric acid, chlorines, bromines, fluorides, iron, chromium/Cr VI, lead, copper and mercury) are quantified in TW and PAW with test strips (Merck Millipore pH paper 1-14 and 16 in 1 water test strips).

Colorimetric assays are performed to measure nitrite and hydrogen peroxide concentrations. In a well of 96-well plate, 10 µL of PAW are diluted in 190 µL of TW, reacting with 50 µL of Griess reagent (Sigma Aldrich, Saint-Quentin Fallavier, France) during 10 min for nitrite quantification. In another well, 200 µL of PAW are reacting with 50 µL of Titanium Oxysulfate IV (TiOSO$_4$ from Sigma-Aldrich, Saint-Quentin Fallavier, France) for hydrogen peroxide quantification. Absorptions are recorded by a Biotek Cytation 3 spectrophotometer, at 548 nm for nitrite and 409 nm for hydrogen peroxide.

## 2.7. Statistical analysis

Descriptive statistics, including means and standard deviations (STD) are calculated for each group to summarize the data. To determine if there are statistically significant differences between the groups, a one-way analysis of variance (ANOVA) is performed. The analysis is conducted at three significance levels: 0.05, 0.01, and 0.001. Group means are compared to evaluate variance within and between groups, and p-values at each significance level are used to determine the presence of significant differences.

# 3. Results

In this section, we first characterize the physicochemical properties of the plasma phase (properties associated with the DAP approach) and then of plasma-activated water (properties associated with the PAW approach). Next, we examine how these approaches can impact the germination parameters of maize as well as the initial stages of seedling development.

## 3.1. Plasma characterization of the DBD for dry atmospheric plasma priming

While different gases can be utilized to treat seeds in a dielectric barrier device, it is always economically more relevant to utilize ambient air to generate cold plasma. This general statement is however limited to seeds with typical diameters of 4 mm maximum (carrots, tomatoes, pepper, …) otherwise gas breakdown cannot occur. Since in our work, maize seeds present a typical diameter of 7.2 ± 0.5 mm, the DBD is supplied in helium and oxygen, forming a gas mixture in which electrical energy is deposited by the high voltage generator.

This configuration results in a plasma voltage of sinusoidal waveform, with an amplitude of 8 kV at 250 Hz, as plotted in **Fig. 2a**. The current profile exhibits a periodic capacitive component as well as a conductive component characterized by a stochastic distribution of current peaks that alternate in polarity with the sign of $V_{plasma}$. In this plasma source, the dielectric barrier accumulates electrical charge, resulting in a delay in current flow relative to the voltage, typically observed as a 90° phase shift where the current lags behind the voltage peaks. Electrical power deposited in the gaseous phase is measured following Lissajous method by interconnecting a non-perturbative capacitor (10 nF) between counter-electrode and ground. Hypothesizing electrical charge conservation along the whole electrical circuit, $Q_{plasma}$ is plotted *versus* $V_{plasma}$ in **Fig. 2b**. A value of 4.1 W is found, which, taking into account the volume of the DBD reactor (80 cm$^3$) and the volume occupied by the seeds (30 cm$^3$), represents a power density of approximately 4.1/50 = 82 mW/cm$^3$.

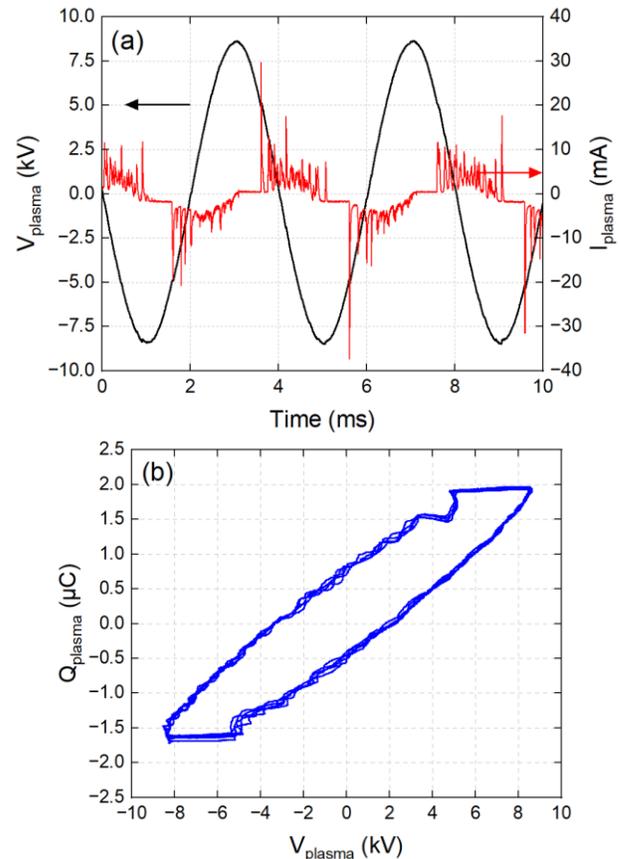

*Figure 2. Electrical characterization of the plasma phase (a) Temporal profiles of plasma voltage and current, (b) Lissajous curve associated to electrical signals from (a).*

*Table 2. Identification of the species observed by OES in the plasma phase generated by DBD for DAP - SPS: second positive system, FNS: first negative system.* [36], [37]

| Species | λ (nm) | Lower level | Upper level | System |
|---|---|---|---|---|
| NO* | 248.8 - 249.3 | B$^2$Π | X$^2$Π | B |
| OH* | 308.90 (head) | A$^2$Σ$^+$ | X$^2$Π | Angstr. |
| N$_2$* | 337.13 (head) | C$^3$Π$_u$ | B$^3$Π$_g$ | SPS |
| N$_2^+$ | 391.44 (head) | B$^2$Σ$^+_u$ | X$^2$Σ$^+_g$ | FNS |
| He I | 706.52 | 1s2p ($^3$P°) | 1s3s ($^3$S) | - |
| O I | 777.42 | 2s$^2$2p$^3$($^4$S°)3s ($^5$S°) | 2s$^2$2p$^3$($^4$S°)3p ($^5$P) | - |

The active species from the plasma phase are analyzed by OES, over a 200-800 nm range and the corresponding transitions are listed in **Table 2**. The optical spectrum in **Fig. 3a** is mostly dominated by the gamma system of nitrite oxide (NO*) and the second positive system of molecular nitrogen ($N_2$*). Unsurprisingly, the spectrum contains the atomic line of helium (He*) at 706.5 nm (**Fig. 3f**) and, in addition to $N_2$* for instance headed at 337.1 nm in **Fig. 3c**, a band of $N_2^+$ ions, headed at 391.4 nm (**Fig. 3d**). Among the short lifespan reactive species, several oxygenated radicals are identified, namely NO* radicals double-headed at 248.8 nm and 249.3 nm (**Fig. 3b**) but also OH* radicals at 308.9 nm (**Fig. 3e**) and O* radicals at 777.4 nm (**Fig. 3g**). These radicals are expected to induce biological effects, as further detailed in the Section 3.4. and Section 4.

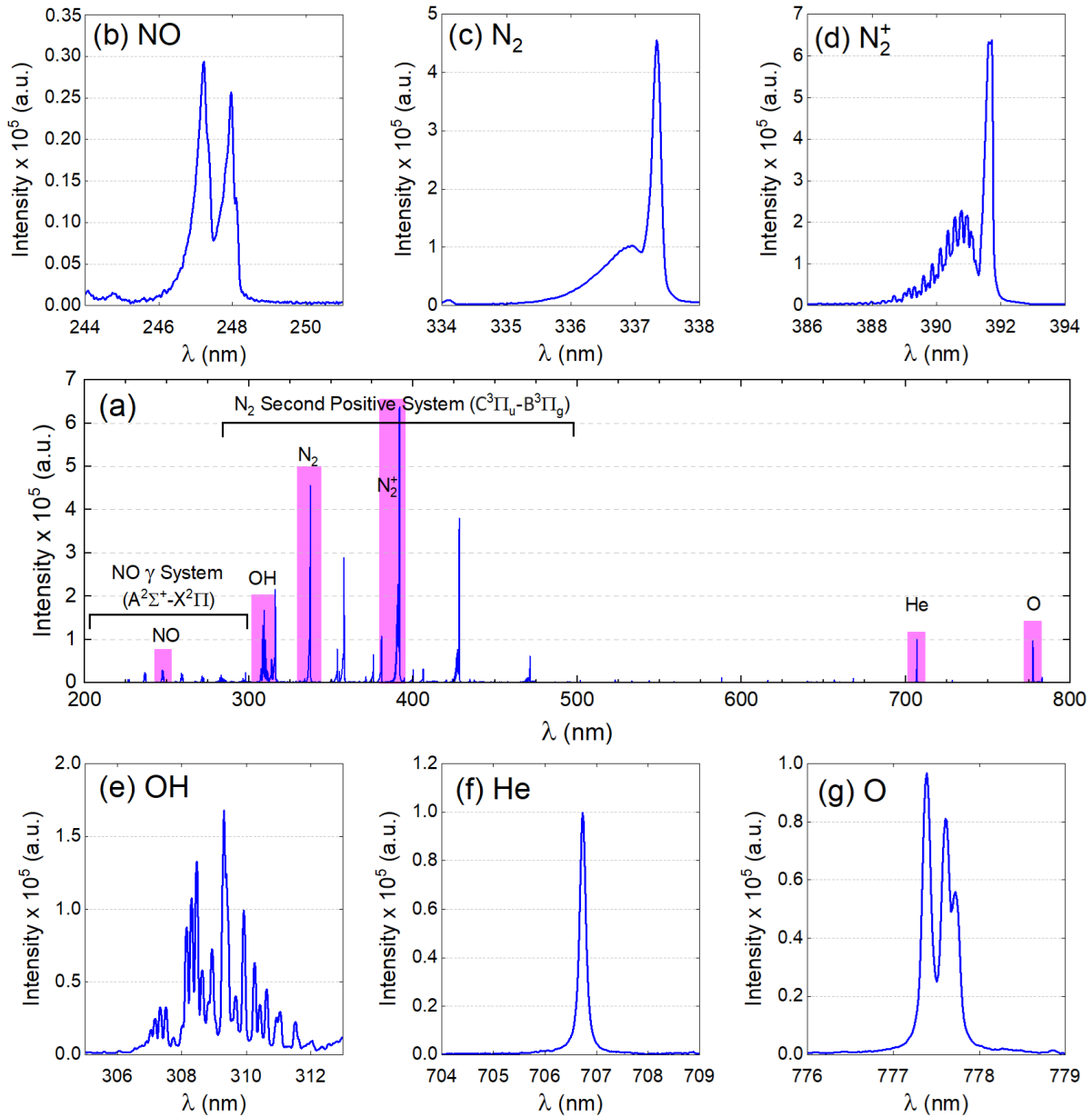

*Figure 3. Optical Emission Spectroscopy of the He-air plasma phase (a) Full spectrum from 200 to 800 nm, (b) NO* radicals double-headed at 248.8 nm and 249.3 nm, (c) $N_2$* headed at 337.1 nm, (d) $N_2^+$ ions headed at 391.4 nm, (e) OH* radicals at 308.9 nm, (f) He* headed at 706.5 nm and (g) O* radicals at 777.4 nm.*

In addition to their electrical and chemical properties, cold plasmas also exhibit radiative properties that may influence seeds properties, either beneficial or detrimental. To investigate this hypothesis, the optical spectrum of He-air plasma is plotted over the 200-800 nm range (**Fig. 4a**) and compared with the optical spectrum of a cloudy sky in Paris (April 2024), serving as an arbitrary yet relevant reference. **Fig. 4a** demonstrates that the overall irradiance of He-air plasma is significantly lower than natural light, with the exception of peaks at 337.1 nm and 391.4 nm. Each of these spectra can be divided into 4 spectral regions:

UVC (100-280 nm), UVB (280-315 nm), UVA (315-400 nm) and Visible/Near InfraRed (400-800 nm), also labeled Vis/NIR. For each of these regions, the area under the spectrum corresponds to the total energy emitted by the plasma source; it can be reported either on absolute scale (**Fig. 4b**) or relative scale (**Fig. 4c** and **4d**).

These two figures clearly highlight that regardless of the spectral ranges, the total radiation energy emitted by the plasma ($1.2 \times 10^8$ a.u.) is almost 100 times lower than that emitted by the cloudy sky ($1.1 \times 10^{10}$ a.u.). Therefore, even though the plasma source emits 6% of its energy in the UVC range and 55.3% in the Vis/NIR range, these components are negligible compared to those of ambient light (cloudy sky). Since seeds are insensitive to ambient light during the germination process, it is reasonable to consider that much weaker radiative properties of the He-air plasma are unlikely to induce biological effects, whether beneficial or detrimental.

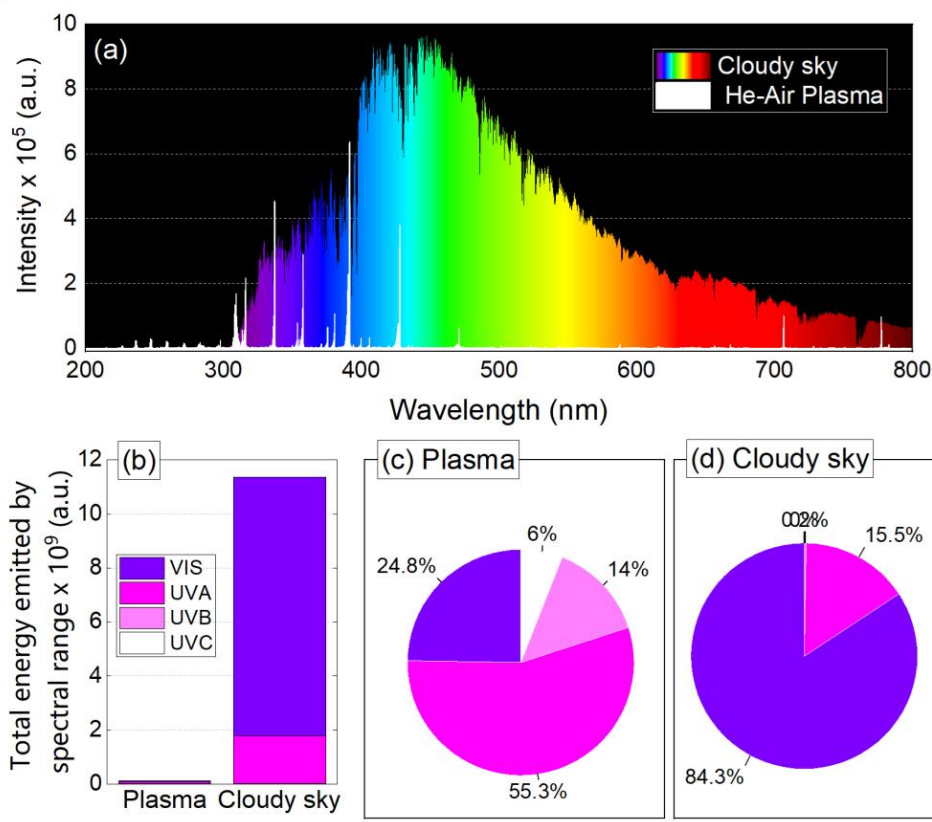

*Figure 4. (a) Optical emission spectra of two radiative sources: the He-air plasma and the ambient light corresponding to a cloudy sky, (b) Total energy emitted by these two sources in absolute scales, (c-d) Distribution of emitted energy by spectral range for each source, classified into four spectral ranges: UVC (100-280 nm), UVB (280-315 nm), UVA (315-400 nm) as well as visible/Near Infrared (400-800 nm).*

## 3.2. Plasma-activation of water: characterization of the plasma phase

For the activation of tap water, the SPED is utilized to generate a DC discharge in ambient air, between the single hollow brass cylinder and the swirling tap water, as sketched in **Fig. 1b**. This cold plasma process operates at a voltage of 4200 V and a current of 30 mA, resulting in an electrical power deposition of 126 W: a value 30 times higher than power deposited on seeds using the DBD source for DAP. The active species of the gaseous phase are characterized by OES, over a 200-800 nm range. As depicted in **Fig. 5a** and detailed in **Table 3**, the predominant emissions are from OH* radicals around 306 to 322 nm (**Fig. 5b**) and from the second positive system of molecular nitrogen ($N_2$*), most notably at 337.1 nm (**Fig. 5c**).

Furthermore, the spectrum contains the highly emissive lines of copper (Cu I) at 324.8 nm and 327.4 nm (**Fig. 5d**) and, zinc (Zn I) at 472.2 nm and 481.1 nm (**Fig. 5e**). The presence of these copper and zinc lines can be explained by the composition and erosion of the brass electrode. Brass is an alloy primarily composed of copper and zinc. The intense heating generated by the cold plasma in ambient air causes erosion of the brass electrode, releasing copper and zinc atoms into the plasma phase, which then emit their characteristic spectral lines. It is worth stressing that this erosion remains however very limited since the mass of the HV electrode remains unchanged (scales accuracy: 0.1 mg) even after 1 hour of plasma operation.

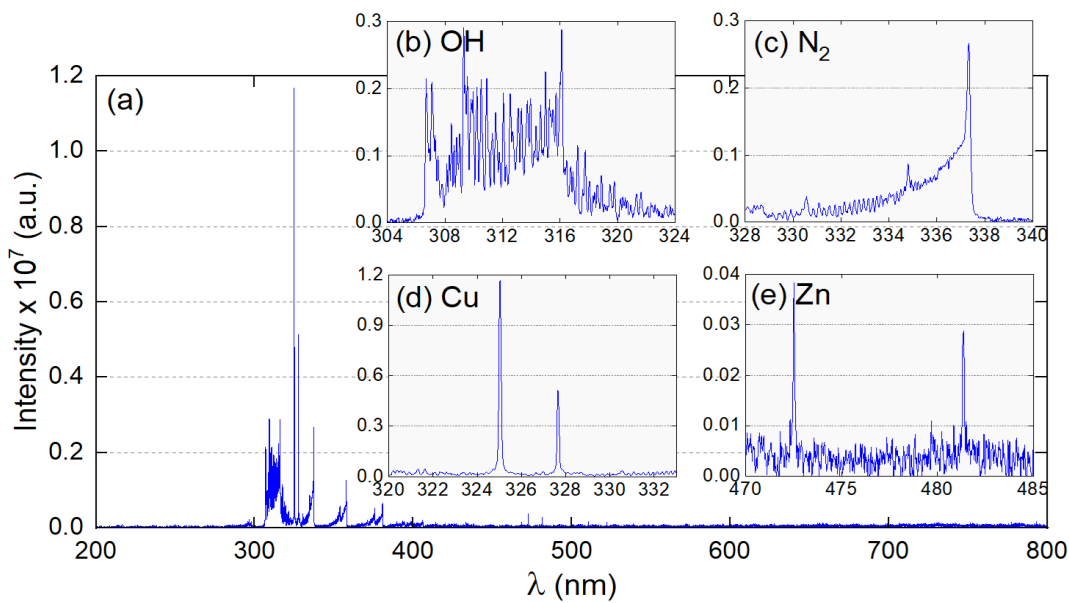

*Figure 5. Optical Emission Spectroscopy of the plasma phase (a) Full spectrum from 200 to 800 nm, (b) OH* radicals from 304 to 324 nm, (c) $N_2$* ($C^3\Pi_u$-$B^3\Pi_g$) headed at 337.1 nm, (d) Cu I ($3d^{10}4s$-$3d^{10}4p$) at 324.8 nm and 327.4 nm, (e) Zn I ($3d^{10}4s4p$-$3d^{10}4s5s$) at 472.2 nm and 481.1 nm.*

*Table 3. Identification of the species observed by OES in the plasma phase generated by SPED for PAW. [36], [37]*

| Species | λ (nm) | Lower level | Upper level | System |
|---|---|---|---|---|
| OH* | 308.90 (head) | $A^2\Sigma^+$ | $X^2\Pi$ | Angström |
| $N_2$* | 337.13 (head) | $C^3\Pi_u$ | $B^3\Pi_g$ | SPS |
| Cu I | 324.75 | $3d^{10}4s$ ($^2S$) | $3d^{10}4p$ ($^2P°$) | - |
| Cu I | 327.39 | $3d^{10}4s$ ($^2S$) | $3d^{10}4p$ ($^2P°$) | - |
| Zn I | 472.21 | $3d^{10}4s4p$ ($^3P°$) | $3d^{10}4s5s$ ($^3S$) | - |
| Zn I | 481.05 | $3d^{10}4s4p$ ($^3P°$) | $3d^{10}4s5s$ ($^3S$) | - |

### 3.3. Plasma-activation of water: characterization of the liquid phase

Unlike the DBD source, which generates radicals such as OH*, O*, and NO*, the DC discharge from the SPED predominantly produces OH* radicals with more intense emissions. The hydroxyl radical is highly reactive and serves as a potent oxidant in both chemical and biological systems [38]. It can react with a broad range of organic and inorganic molecules, playing a crucial role in numerous degradation and oxidation processes. It is essential to analyze the effects of SPED on tap water by measuring its chemical and physical parameters both before and after exposure, as detailed in **Table 4**. For context, 90 mL of tap water is treated for 5 minutes.

*Water chemical properties*
Significant reductions in total alkalinity and carbonates are observed. Total alkalinity, which assesses the water's capacity to neutralize acids, is largely attributed to bicarbonates, carbonates, and occasionally hydroxides, shows a substantial reduction from 140 mg/L to 60 mg/L. This significant drop suggests that the plasma treatment may degrade or transform these buffering agents. The decline in carbonate concentration from 280 mg/L to 60 mg/L supports this view, indicating that carbonate ions may interact with plasma. The reactive species produced by the plasma within the PAW could convert carbonate ions into carbon dioxide and water, thus reducing both alkalinity and carbonate levels. This alteration can diminish the water's ability to buffer pH fluctuations, thereby increasing its vulnerability to acid or base additions.

Water hardness, which is primarily determined by the concentrations of calcium and magnesium ions, remains constant before/after SPED treatment, with a value close to 250 mg/L. This suggests that these ions are not significantly reacting or precipitating out of solution during the plasma treatment and cannot be responsible for beneficial or detrimental effects on the growth of seedlings.

The concentration of chlorine ($Cl_2$), whether free or total, remains nearly zero (less than 0.5 mg/L) both before and after plasma treatment. In the PAW, chlorine is negligible and is unlikely to have any biological effects. Chlorine must not be confused with concentration of chlorides (often measured as chloride ions, $Cl^-$) whose value is 20 mg/L, as reported in **Table 1**.

The concentrations of nitrate and nitrite ions increase dramatically, from 25 mg/L to 300 mg/L and from 0.03 mg/L to 56.11 mg/L respectively. This important change is attributed to the plasma treatment's ability to convert molecular nitrogen ($N_2$), which is naturally present in both ambient air and dissolved in water, into more reactive forms such as nitrates ($NO_3^-$) and nitrites ($NO_2^-$). This conversion occurs through collisional interactions between $N_2$ and either oxygen molecules or water vapor, facilitating a process known as nitrogen fixation.

The concentration of hydrogen peroxide increases substantially (from 0.26 mg/L to 18.48 mg/L), as already reported in the literature [39].

No detectable changes in bromine, cyanuric acid and fluoride are detected before/after plasma treatment. Same conclusion applies for metals (iron, chromium, lead, copper, mercury), underlying that even if erosion of copper and zinc is detected by OES (**Fig. 5**), the resulting concentrations in the liquid phase are undetectable.

*Water physical properties*
Concerning the physical parameters, only a slight decrease in pH is measured (from 7.8 to 6.8), hence suggesting a limited PAW acidification resulting from the formation of acidic compounds and/or consumption of alkaline substances. Similarly, a slight increase in conductivity is measured from 532 µS/cm to 574 µS/cm). This increase could be attributed to increased ionization or the creation of new ionic species during plasma treatment.

*Table 4. Physicochemical properties of tap water before SPED treatment (Control) and after 5 min of plasma exposure (PAW).*

|  |  | Control | PAW 5 min |
|---|---|---|---|
| **Chemical param.: Concentration (mg/L)** | Total alkalinity | 140 ± 30 | 60 ± 20 |
|  | Carbonate | 280 ± 20 | 60 ± 20 |
|  | Hardness | 250 ± 50 | 250 ± 50 |
|  | Cyanuric acid | 0 ± 20 | 0 ± 20 |
|  | Total chlorine | 0.0 ± 0.5 | 0.5 ± 0.5 |
|  | Free chlorine | 0.0 ± 0.5 | 0.0 ± 0.5 |
|  | Bromine | 0.0 ± 0.5 | 0.0 ± 0.5 |
|  | Nitrate | 25 ± 10 | 300 ± 50 |
|  | Nitrite | 0.03 ± 0.03 | 56.11 ± 15.19 |
|  | Hydrogen peroxide | 0.26 ± 0.06 | 18.48 ± 0.95 |
|  | Metals (Iron - Chromium/Cr (VI) - Lead - Copper - Mercury) | 0.0 ± 0.5 | 0.0 ± 0.5 |
|  | Fluoride | 0.0 ± 0.5 | 0.0 ± 0.5 |
| **Phys. p.** | pH | 7.8 ± 0.4 | 6.8 ± 0.4 |
|  | Conductivity (µS/cm) | 532 ± 8 | 574 ± 13 |

## 3.4. Influence of DAP priming on maize germi-native parameters

Having examined the main physicochemical properties of the DAP and PAW approaches, we now turn our attention to their biological effects on maize seeds and seedlings.

**Fig. 6a** shows the germination rate of corn seeds treated with plasma for different durations (0, 5, 10, 15 and 20 minutes) over a period of up to 16 days. Seeds treated for 0 minute (control group) show a germination rate of only 65 % while it is higher than 90 % for plasma exposure times longer than 5 minutes. Seeds treated for 20 minutes reach a higher germination rate more quickly than those treated for shorter periods.

Median germination times for the 5 groups, summarized as a bar chart in **Fig. 6b**, decrease as plasma treatment time increases. Untreated seeds take the longest to germinate (about 8 days), whereas those treated for 20 minutes germinate in 5 days, which represents a kinetic gain of (8-5)/8 = 37.5 %. It is also worth noting that no significant difference is observed between 15 min and 20 min, which is consistent with previous works we performed on lentil seeds, in which we demonstrated that treatment times as long as 90 min only slightly improve the germination parameters [40]. Overall, these results indicate that DAP priming can increase germination rate of maize seeds in a shorter time, hence opening the way for sustainable agricultural practices (crop yields optimization, planting schedules, ...).

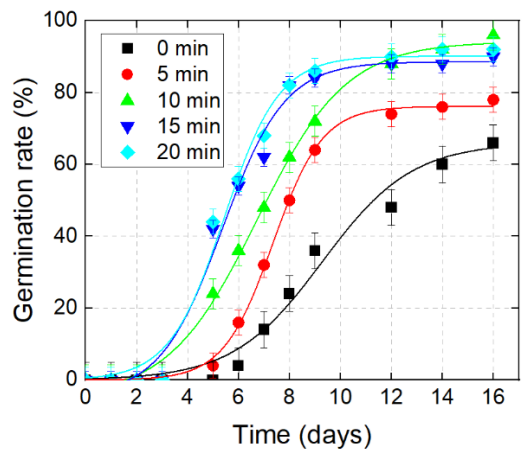

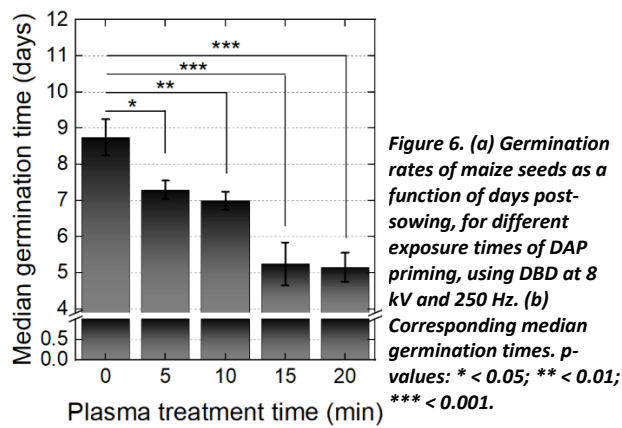

*Figure 6. (a) Germination rates of maize seeds as a function of days post-sowing, for different exposure times of DAP priming, using DBD at 8 kV and 250 Hz. (b) Corresponding median germination times. p-values: * < 0.05; ** < 0.01; *** < 0.001.*

## 3.5. Combining pre- and post-germinative plasma treatments to improve seedlings growth parameters

The experiment continues over a three-week period to explore the effects of DAP on seedling growth, in comparison or in combination with PAW. For this, a triplicate experiment is carried out in which we consider four groups of 24 seeds each, with seedling irrigation on days 16, 22, 26 and 33, using either TW or PAW. The configuration of the four groups is as follows:
- Control group (seeds not primed and seedlings irrigated with TW)

- DAP group (seeds primed in DBD for 20 min and seedlings irrigated with TW)
- PAW group (seeds not primed and seedlings irrigated with PAW for 5 min)
- DAP+PAW group (seeds primed in DBD for 20 min and seedlings irrigated with PAW for 5 min)

The impact of these plasma treatments on the early growth stages of maize seedlings is depicted in **Fig.7** which corresponds to a photograph taken the 36$^{th}$ day after seed's imbibition. Two additional photographs are introduced in Appendix to corroborate the reliability of this study (**Fig. 12** and **Fig. 13**). Qualitatively, **Fig. 7** clearly demonstrates that maize seedlings treated with plasma (DAP, PAW, DAP+PAW) are more robust and taller than those in the control group. Notably, seedlings irrigated with PAW exhibit better growth than those irrigated with tap water. Additionally, the DAP+PAW combination results in the most significant growth enhancement, suggesting a synergistic effect when both treatments are combined.

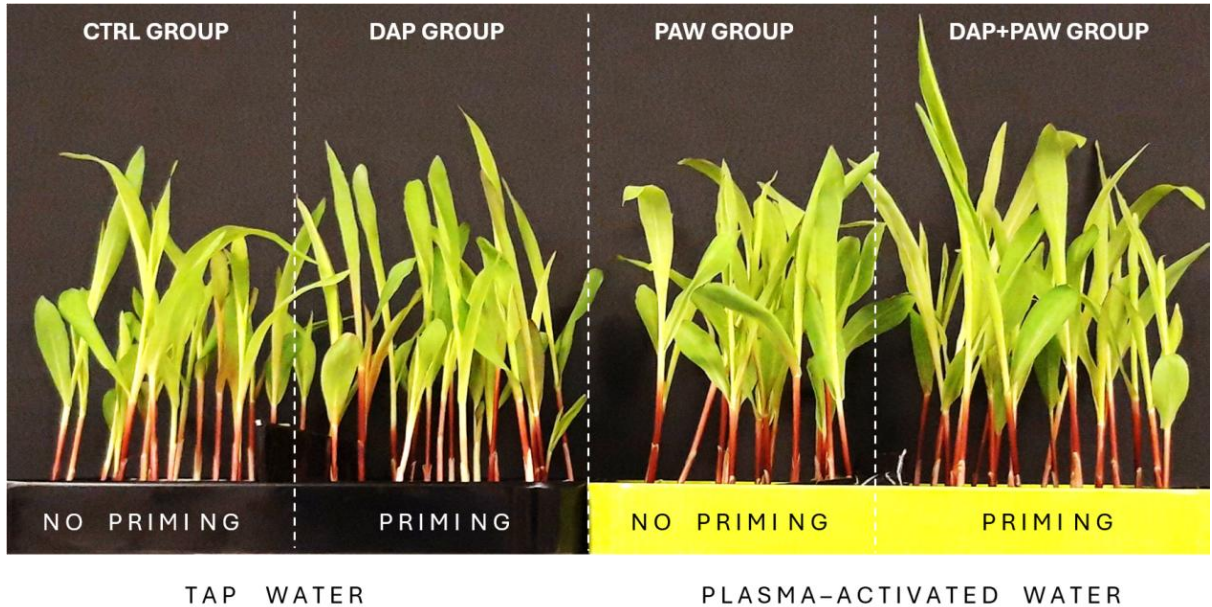

*Figure 7. Photograph of maize seedlings on the 36$^{th}$ day after been maintained at 15°C and exposed to light 6 hours per day for the following groups: CTRL, seeds unexposed to DAP nor PAW; DAP, seeds exposed to DAP and irrigated by tap water; PAW, seeds unprimed by DAP but irrigated with PAW; and DAP priming + PAW irrigation, combining both plasma processes.*

A more analytical approach is carried out throughout the experiment, by measuring several seedlings growth parameters, namely stem length, hypocotyl length, number of leaves and collar diameter. The monitoring of stem lengths over a 2-weeks period, as reported in **Fig. 8**, drives to notable differences:
- Seeds primed by DBD for 20 minutes (represented by red and blue bars) typically result in longer seedlings compared to their non-primed counterparts (black and green bars). For example at Day 36, a stem length of 16.1 cm is observed for the DAP group vs only 14.9 cm for CTRL group.
- Seeds irrigated with PAW (green and blue bars) exhibit improved growth relative to those irrigated with tap water (black and red bars) only from Day 30. Before, stem lengths from CTRL group (e.g. 4 cm at Day 24) are always longer than those from PAW group (e.g. 2.8 cm at Day 24).
- DAP priming is more effective than PAW irrigation for follow-up periods prior to day 30 (for example, stems at day 22 are twice as long in the DAP group than in the PAW group).
- The most pronounced growth is seen in the group that receives both DAP priming and PAW irrigation (blue bars), suggesting a synergistic effect of the combined treatments on seedling development.

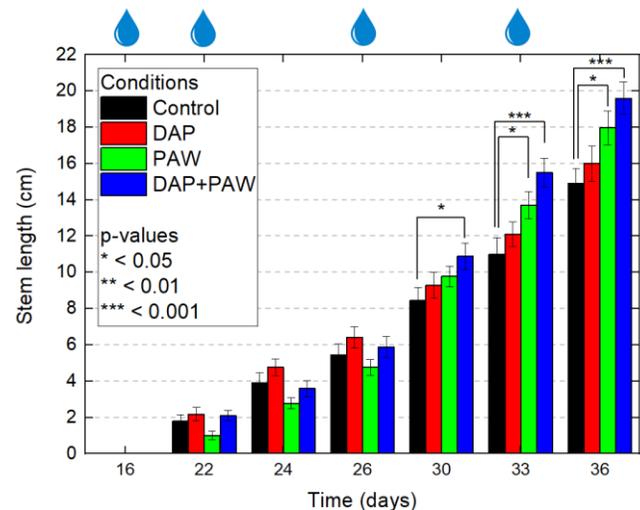

*Figure 8. Stem length of maize seeds as a function of post-sowing days for the current conditions: CTRL, DAP, PAW and DAP+PAW groups, as presented in Figure 7. Water droplet symbols indicate the days on which seeds are irrigated with either tap water (TW) or plasma-activated water (PAW).*

For the sake of clarity, hypocotyl length, number of leaves and collar diameter are reported in **Fig. 9** on the 36th day. It turns out that (i) PAW treatment significantly improves hypocotyl length and leaf number compared with DAP priming alone, (ii) the combined DAP+PAW treatment shows the greatest improvement in these parameters. Interestingly, collar diameter seems not affected by any plasma treatment, maintaining a constant value close to 3.5 mm in all groups.

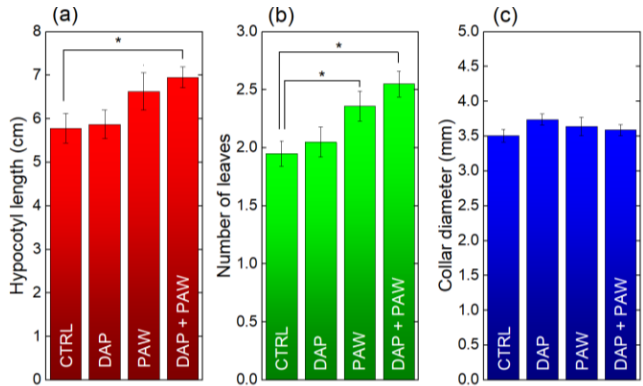

*Figure 9. Seedlings growth parameters measured on the 36th day: (a) hypocotyl length, (b) number of leaves and (c) collar diameter of stems. Conditions related to CTRL, DAP, PAW and DAP+PAW groups are identical to those in Figure 7.*

In addition to seedling growth parameters, the relative fresh masses of seedlings are reported in **Fig. 10** from 41$^{st}$ day (seedlings removed from soil) to 51$^{st}$ day (end of monitoring). Relative fresh mass provides an indirect measurement of water content as it can reveal whether the seedlings are experiencing stress or adapting to drought conditions. **Fig. 10** shows the relative fresh masses of seedlings. For each group, 10 seedlings are isolated for this assessment. As the mass of these 10 seedlings can vary from one group to another, the fresh masses are presented in relative terms, with the 41$^{st}$ day serving as the baseline. The data show that:

- PAW treatment alone does not alter the fresh mass of seedlings while DAP priming increases the value of this parameter by around 5% at 42$^{nd}$, 43$^{rd}$ and 45$^{th}$ day before becoming negligible thereafter. This means that DAP priming can optimize seedling growth parameters, by promoting two factors: their ability to retain moisture, and potentially by increasing nutrient and energy use efficiency.
- The combined DAP+PAW treatment induces a similar behavior to that of the DAP treatment.
- All relative fresh masses tend towards a relative dry mass value of 11.5% at day 51. This means that, although plasma treatments can influence hydration and initial vigour, they have no long-term or significant impact on seedlings' dry matter (dry biomass). This result indicates that the effects of plasma treatments are more pronounced in the early stages of growth or in environments where water management is critical.

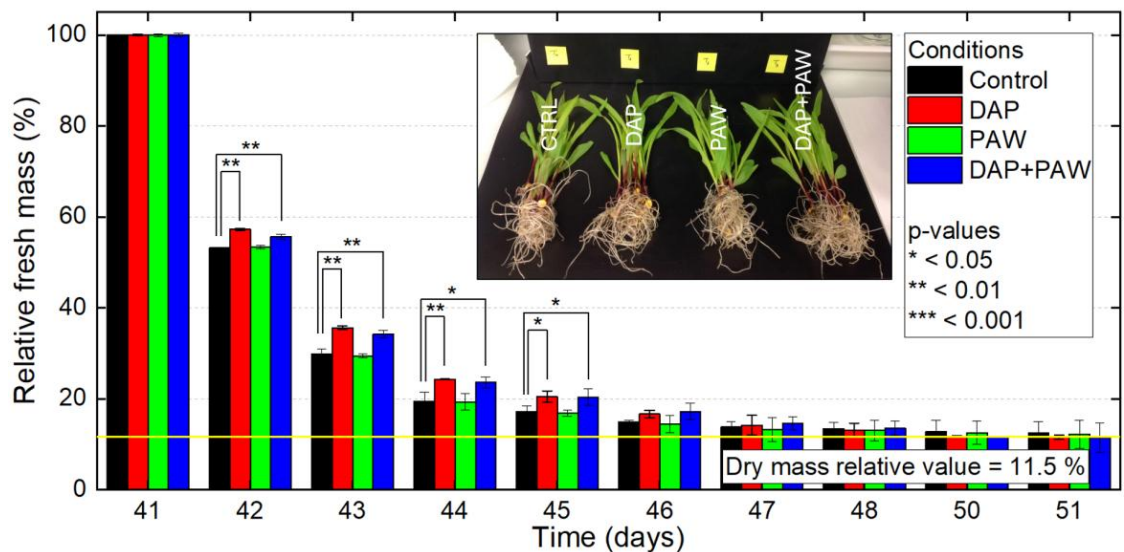

*Figure 10. Last day fresh masses of the seedlings as a function of the drying time, converging to the dry mass relative value of 11.5% for the current conditions presented in Figure 7.*

## 3.6. Pre- and post-germinative plasma treatments reduce seedlings photosynthesis and stomatal density

In the previous section, seedling growth parameters (hypocotyls, leaves and collar diameter) indicate that all plasma treatments significantly enhance seedling growth dynamics, with effects maximized by combining DAP and PAW. To investigate whether these macroscopic enhancements in growth are associated with physiological and structural changes at the cellular level, further studies are carried out on maize seedling leaves: (i) measurements of chlorophyll content to assess changes in photosynthetic activity and (ii) auto-fluorescence imaging to detect microstructural alterations.

Chlorophyll content is measured in the leaves of 10 seedlings from each of the four groups, leading to the results shown in **Fig. 11a**. While the seedlings in the CTRL group exhibit a chlorophyll concentration close to 115.3 µmol/m$^2$, the DAP and PAW treatments have values of only 106.9 µmol/m$^2$ and 89.8 µmol/m$^2$, respectively. When the two treatments are combined, chlorophyll concentration falls further to 91.9 µmol/m$^2$. Consequently, despite the initial positive effects of plasma treatments on leaf number, stem size and hypocotyl length, leaf chlorophyll content does not reflect this growth, but instead shows a slight decrease.

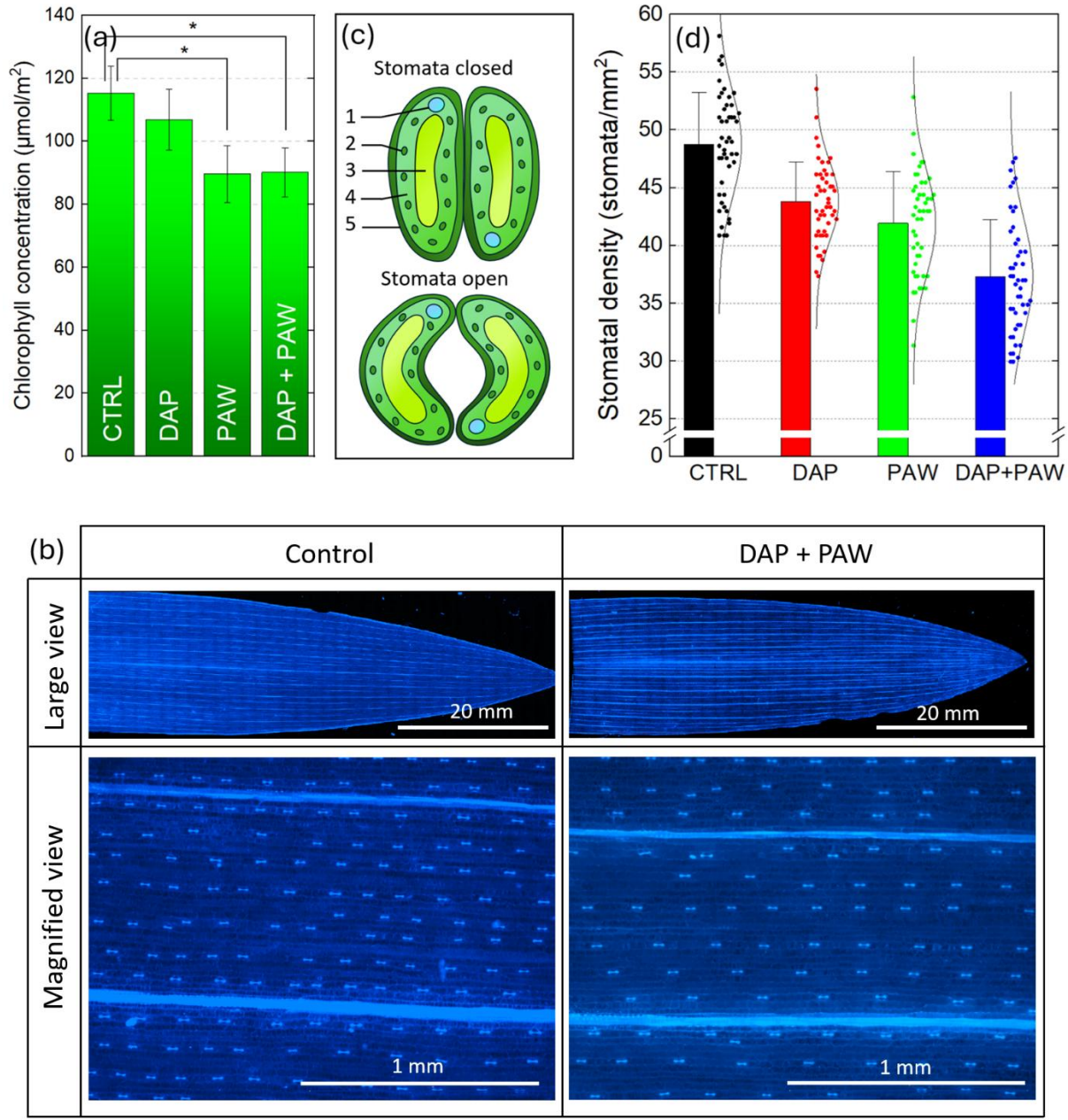

*Figure 11. (a) Chlorophyll concentration in maize leaves for CTRL, DAP, PAW and DAP+PAW groups. (b) Comparative auto-fluorescence imaging of maize seedling leaves from CTRL and DAP+PAW groups, observed on the 36$^{th}$ day (×4 magnification, excitation at 358 nm and emission around 461 nm). (c) Diagram illustrating the structural differences between closed (top) and open (bottom) stomata. Key features labeled are: 1 - Nucleus, 2 - Chloroplasts, 3 - Vacuole, 4 - Guard cell, 5 - Cell wall. (d) Stomatal density measured on maize leaves for CTRL, DAP, PAW and DAP+PAW groups at Day 36. For each group, the statistic is achieved on 50 pictures in magnified views.*

Alongside physiological changes, we have also investigated potential structural alterations at the cellular level. The leaves of maize seedlings are observed using a fluorescence cell imaging device providing a 4x magnification, in DAPI mode but without adding any DAPI agent so that only auto-fluorescence is detected. Microphotographs in **Fig. 11b** show the upper surface of selected leaves presenting the same macroscopic dimensions, considering only CTRL and DAP+PAW groups for clarity. These microphotographs (specifically those with magnified views) reveal regular patterns of paired nuclei, spaced approximately 90-130 μm apart. These paired nuclei are associated with the two guard cells of each stoma, as illustrated in **Fig. 11c**, regardless of whether they are open or closed. This auto-fluorescence facilitates the visualization of stomata, which are vital for photosynthesis as they manage gas exchange ($CO_2$ intake and $O_2$ release) and water regulation (balancing water vapor loss through transpiration with $CO_2$ intake necessary for photosynthesis). In each group, the number of stomata is counted over 50 magnified views taken from different leaves. **Fig. 11d** shows that the CTRL group exhibits a stomatal density of about 48 stomata/mm$^2$, while DAP and PAW treatments result in lower densities: 44 and 42 stomata/mm$^2$, respectively. Interestingly, their combination further reduces the density to 37 stomata/mm$^2$.

It is worth stressing that our fluorescence cell imaging device is limited in counting the number of stomata and does not permit further characterizations. This constraint implies that the observed reduction in stomatal density following plasma treatment could potentially be offset by an increase in stomatal activity compared to the CTRL group.

## 4. Discussion

Our study builds on the work of Sivachandiran *et al*. and Rashid *et al*., who were the first to highlight the synergistic effects between DAP and PAW. Our article reinforces this message and extends it to a large-scale agronomic crop model: maize. It addresses the challenges posed by the treatment of large-diameter seeds with DBD in ambient air. In our case, we chose to add a noble gas, helium, as an alternative to working at low pressure. In addition, we chose to hydrate the seedlings with tap water rather than demineralized water. This decision was taken partly because it is more practical in applications, and partly because demineralized water, whether plasma-treated or not, inevitably comes into contact with the substrate (soil, coconut fibre, etc.) before and during its interaction with the seed. In any case, understanding the direct impact of water molecules on seeds is compromised as long as sowing is achieved in a substrate.

We have demonstrated that the DAP treatment significantly increases the germination rate of maize seeds, with values as high as 92%, compared to 65% for the control group. This increase is accompanied by a median germination time that is 37.5% lower than control group, hence suggesting that DAP priming not only enhances germination rate but also accelerates the process. Second, we have demonstrated that the combined DAP and PAW treatments produce the most significant improvements in seedling growth. This synergistic effect is the result of two components: DAP priming improves initial germination and early seedling vigor, while PAW treatment improves overall growth. The increase in seedlings growth parameters such as hypocotyl length and leaf number underlines the potential of plasma treatments to optimize crop growth.

Additionally, physiological analyses in Section 3.4. reveal that while plasma treatments enhance growth parameters, they also lead to a reduction in chlorophyll content and stomatal density in the leaves. These findings let us hypothesize that ROS delivered by plasma via the dry (DAP) or liquid (PAW) approach may promote seedling growth via mechanisms that render the photosynthetic process secondary. As outlined afterwards, these mechanisms may include signal transduction pathways, hormonal interactions, modulation of gene expression, response to oxidative stress and metabolic changes.

ROS generated by cold plasma can act as signaling molecules that trigger various signal transduction pathways involved in growth regulation. They can activate cascades of mitogen-activated protein kinases (MAPKs) which play a crucial role in cell division and elongation [41]. Adhikari *et al*. have demonstrated that PAW treatment leads to activation of MAPK signaling pathways in tomato seedlings, which are known to play a crucial role in regulating plant defense responses and growth processes [42]. Complementarily, the works of Li *et al* indicate that in the case of dry atmospheric plasma treatment of tomato seeds, the expression of MAPK1 and MAPK2 is significantly increased, along with RBOH1 and hydrogen peroxide ($H_2O_2$) concentrations [43]. Even if, to the best of our knowledge, the literature does not provide such mechanisms in the case of maize treated by plasma sources, signal transduction pathways appear as a reasonable assumption to promote seedlings growth while maintaining chlorophyll content and stomata density at reduced levels.

ROS could also interact with plant hormones such as auxins, gibberellins and cytokinins, thereby enhancing cell division and elongation and promoting rapid growth. Cytokinins are known to directly promote chlorophyll synthesis by activating relevant biosynthetic pathways and by delaying leaf senescence [44]. Cytokinins can also have a direct role in regulating stomatal formation by influencing the differentiation of stomatal precursor cells, leading to the formation of new stomata [45]. Since such effects are not obtained in our work, this suggests that DAP and PAW could not promote significant production of cytokines, or could even decrease them. This scenario has already been demonstrated for maize seeds exposed less than 3 min to gliding arc plasma treatment [46]. Gibberellins indirectly influence chlorophyll synthesis by promoting overall seedlings growth, which increases the demand for chlorophyll as the plant produces more leaves and expands its photosynthetic capacity. However, gibberellins do not directly regulate the pathways for chlorophyll synthesis, they can only affect the expression of genes related to chloroplast development and chlorophyll biosynthesis. An increase in gibberellin levels seems therefore appropriate in our experimental conditions; as already been demonstrated on other models like sunflower [47], radish [48] and Arabidopsis thaliana [49].

In addition to signal transduction pathways and hormonal interactions, ROS from the plasma phase could also participate to mechanisms which improve growth parameters but reduce chlorophyll content and stomatal density:
- The modulation of gene expression by plasma-generated ROS, thus driving the expression of growth-

associated genes, as proposed by Iranbaksh *et al*. [50]. ROS could upregulate genes involved in cell cycle progression and downregulate those involved in chlorophyll biosynthesis and stomatal development. In addition, cold plasma could induce epigenetic modifications that affect gene expression in seedlings [51].
- The mild oxidative stress of seedlings, leading to the activation of stress response pathways, such as more resources allocated to produce antioxidants and stress-related proteins. In turn, integrity at cell level could be maintained during rapid development.
- Metabolic shifts, i.e. changes in seedlings' primary and secondary metabolism [52]. To sustain growth under stressful conditions, seedlings could produce more secondary metabolites (e.g. antioxidants) while simultaneously diverting resources from chlorophyll synthesis.

## 5. Conclusion

This study demonstrates the significant impact of two cold plasma treatments on maize seeds' germination and seedlings' development. DAP treatment enhances the germination rate (from 65% to more than 90%) while reducing the median germination time by 37.5%. These results indicate that DAP priming not only enhances overall germination efficiency but also accelerates the process, making it advantageous for agricultural practices that need rapid seedling establishment.

Furthermore, the combination of DAP priming and PAW treatment produces the most significant improvements in seedling growth parameters, including stem length, hypocotyl length and leaf number. This synergistic effect indicates that DAP priming promotes initial germination and vigor, while PAW treatment further stimulates growth, resulting in stronger seedlings, which could optimize crop development and yields.

Physiological analyses reveal that while plasma treatments improve growth parameters, they also lead to a reduction in chlorophyll content and stomatal density in leaves. This suggests that plasma-generated ROS could promote growth via mechanisms that deprive photosynthetic efficiency. These mechanisms may include signal transduction pathways, hormonal interactions, modulation of gene expression, response to oxidative stress and metabolic changes.

Future research should explore the long-term effects of these treatments and their applicability to other crops in order to fully exploit their potential in sustainable agriculture. In this context, it would be valuable to study a broad range of seeds treated with the same DAP and PAW processes, to identify which species exhibit the strongest synergistic effects and to uncover the underlying reasons through biochemical, nutritional, and antioxidant enzyme analyses.

## 6. Acknowledgements

The authors would like to thank the French Embassy in Cameroon for awarding 25 mobility grants to Dr Madeleine TCHUINTE, Minister of Scientific Research and Innovation (MINRESI). The authors also acknowledge Pr. Samuel LAMINSI for useful discussions during the thesis of J. P. KAMSEU MOGO in Cameroon.

## 7. Data availability statement

The data that support the findings of this study are available upon reasonable request from the authors.

## 8. Informed consent statement

Informed consent was obtained from all subjects involved in the study.

## 9. Conflict of interest

The authors declare no conflict of interest.

# 11. Appendix

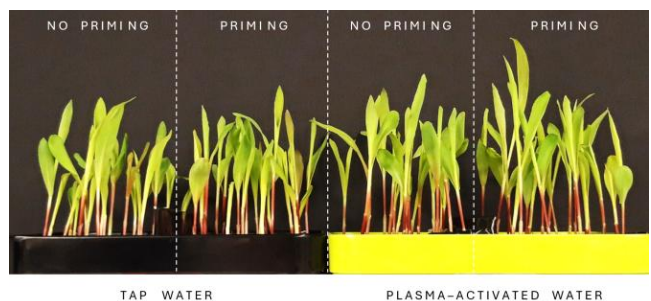

*Figure 12. Photograph of maize seedlings on the 33rd day,* **i.e.** *3 days before the photograph shown in Figure 7. Seedlings are organized into 4 groups: CTRL (seeds without DAP priming and only irrigated with tap water); DAP (seeds primed by the He-air plasma for 20 min at 8 kV and 500 Hz, followed by irrigation with tap water); PAW (seeds without DAP priming but irrigated with plasma-activated water after a 5-minute SPED treatment; DAP+PAW (combination of DAP priming and PAW irrigation).*

*The seedlings are maintained in controlled conditions at 15 °C, with a photoperiod of 6 hours of light and 18 hours of darkness daily.*

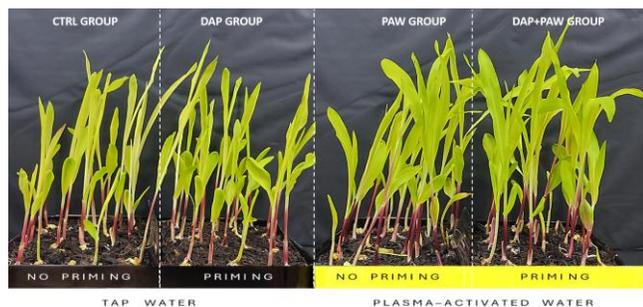

*Figure 13. Independent experimental replicate of the first experiment presented in Figures 7 and 12. Here, the DAP treatment is identical, but the water activation is conducted for 20 minutes instead of 5 minutes. This photo is taken 30 days after the seed imbibition.*